# Capture of field stars by giant interstellar clouds: the formation of moving stellar groups

C. A. Olano★†

*Facultad de Ciencias Astronómicas y Geofísicas, Universidad Nacional de La Plata, Paseo del Bosque, 1900 La Plata, Argentina*

**ABSTRACT**

In the solar neighbourhood, there are moving groups of stars with similar ages and others of stars with heterogeneous ages as the field stars. To explain these facts, we have constructed a simple model of three phases. Phase A: a giant interstellar cloud is uniformly accelerated (or decelerated) with respect to the field stars during a relatively short period of time (10 Myr) and the cloud's mass is uniformly increased. As a result, a number of passing field stars is gravitationally captured by the cloud at the end of this phase; phase B: the acceleration (or deceleration) and mass accretion of the cloud cease. The star formation spreads throughout the cloud, giving origin to stellar groups of similar ages; and phase C: the cloud loses all its gaseous component at a constant rate and in parallel is uniformly decelerated (or accelerated) until reaching the initial velocity of phase A (case 1) or the velocity of the gas cloud remains constant (case 2). Both cases give equivalent results. The system equations for the star motions governed by a time-dependent gravitational potential of the giant cloud and referred to a coordinate system comoving with the cloud have been solved analytically. We have assumed a homogeneous spheroidal cloud of fixed semimajor axis $a = 300$ pc and of an initial density of 7 atoms cm$^{-3}$, with a density increment of 100 per cent and a cloud's velocity variation of 30 km s$^{-1}$, from the beginning to the end of phase A. The result is that about 4 per cent of the field stars that are passing within the volume of the cloud at the beginning of phase A are captured. The Sun itself could have been captured by the same cloud that originated the moving groups of the solar neighbourhood.

## 1 INTRODUCTION

An important problem of Galactic dynamics is the formation of the moving stellar groups, and in particular, of those located in the solar neighbourhood. The existence of kinematic groups of stars, used to be called star streams, was recognized after the pioneering studies of Proctor (1869) and the theory of the *two star streams* of Kapteyn (1905) that was a fundamental step for the development of the Galactic dynamics (for an historical survey on this topic see Antoja et al. 2010). The peculiarity of the moving groups is that even though they are gravitationally unbound systems subject to the tidal disruption, the members of each group have shared a very similar spatial velocity for a long time. An explanation could be that these stellar groups are relics of an old star-forming event (Eggen 1965, 1995, 1996; Asiain, Figueras & Torra 1999) and that they have been tied to the mother gas-cloud, until it has been dispersed (Olano 2001). The existence of groups chemically homogeneous seems to support this hypothesis (De Silva et al. 2007). However, there are other groups that cannot be chemically distinguished from the field stars (Bensby et al. 2007). Another mechanism for forming moving groups involves orbital resonance effects due to the bar and/or the spiral arms of the Galaxy (Dehnen 1998; Fux 2001; Antoja et al. 2011; Quillen et al. 2011).

Certainly, this mechanism can explain the age heterogeneity observed in certain moving groups (De Simone, Wu & Tremaine 2004; Famaey et al. 2005; Famaey, Siebert & Jorissen 2008). Hence, the two models are not incompatible or mutually exclusive, but it would imply that groups originated by two different mechanisms coexist in the solar vicinity.

In this paper, we propose a sole mechanism that could account for both kinds of moving groups: homogeneous and heterogeneous in age. We will demonstrate that the sudden acceleration (or deceleration) of a giant interstellar cloud plus an increment of the cloud mass during the processes is an efficient mechanism to capture field stars passing on the gas cloud. This mechanism favours the capture of stars with velocities similar to the cloud's peculiar velocity produced by the acceleration process. It would explain the formation of moving groups of field stars. The Sun itself would been captured by the cloud that gave origin to the local moving groups (Olano 2001). On the other hand, the formation of stars within the giant gas cloud gave origin to stellar groups of similar ages.

★E-mail: colano@fcaglp.fcaglp.unlp.edu.ar
† Member of The Carrera del Investigador Científico, CONICET, Argentina.

The giant H I clouds (or superclouds), with typical radii of a few hundred parsecs, are observed mainly in and close to the spiral arms of the Galaxy (Elmegreen & Elmegreen 1987). According to the theory of density waves, the gas clouds are comprised and strongly decelerated when they enter the spiral density wave (Roberts 1969, 1970). Another phenomenon that can accelerate gas of huge regions of the Galactic disc and form such giant clouds is the fall of high-velocity clouds (HVCs) on the Galactic disc (see Olano 2004, 2008 and references therein).

Previous studies have been focused on the capture of field stars by molecular clouds (Bhatt 1989) and by collapsing clouds (Whitman, Matese & Whitmire 1991). In both studies, the clouds taken as a whole have been considered at rest with respect to the field stars.

This work is organized in the following manner. In Section 2, we develop a model in which a giant interstellar cloud whose mass and kinematics are subject to abrupt changes is able to capture a percentage of the field stars that are passing through the cloud. The movements of stars under a time-variable gravitational potential of a homogeneous oblate spheroid, representing the cloud, are referred to a coordinate system joined to the cloud which moves with a constant linear acceleration, and the corresponding motion equations are solved analytically (Section 2.1). We derive the velocity distribution of the field stars lying within the cloud at the beginning of the process, providing the initial conditions to derive the orbits of the stars (Section 2.2), and determine the stars that are captured (Section 2.3). In Section 3 and its subsections, we present the results, including the possible capture of the Sun by the same interstellar cloud that originated the local moving groups (Section 3.4) and the formation of moving groups of stars born in the cloud (Section 3.5). Finally, in Section 4, we give the conclusions of this work.

## 2 THE MODEL

The model starts with a gaseous protocloud in which there are field stars passing through the cloud with velocities greater than the escape velocity. Then, the cloud, originally at rest in the local standard of rest (LSR) within the cloud's neighbourhood, is accelerated (or decelerated) during certain time in which the cloud also grows its mass by accretion of the surrounding gas. We will call this period phase A. We will show that at the end of phase A a number of the passing field stars are trapped by the cloud. That is to say that their resulting velocities are less than the escape velocity from the cloud. Then, we consider a stationary period in which the system of the cloud and captured stars is externally not perturbed, but a new stellar population is born within the cloud. We call this period phase B. Finally, the cloud suffers a process of mass-loss with or without deceleration (or acceleration) of the cloud as a whole. In this period, which we call phase C, the cloud loses almost all its gas mass and the captured field stars and those born in the cloud are liberated.

The main simplification of the model is that the cloud and the surrounding field stars are considered as a relatively isolated system, in which the dynamical effects of the general gravitational field of the Galaxy on the capture process can be neglected, and that the external force acting on the gaseous cloud during phase A and phase C does not *directly* affect the stars.

### 2.1 Motion equations and their analytical solutions

To facilitate the analytical treatment, the mass distribution of the cloud is represented by an oblate spheroid, or oblate ellipsoid, of homogeneous gaseous density. The dimensions of the ellipsoid remain constant during all the process, only the density changes with time. Since the gas density is considered uniform within the cloud, this is only a function of the time $t$, that we represent by the expression

$$\rho(t) = \rho_0(1 + \lambda t), \quad (1)$$

where $\rho_0$ is the initial density and $\lambda$ is a constant. We shall study the motions of the stars in the inner region of the cloud (the spheroid) under its gravitational field. Given the cloud's geometry we have adopted, it is convenient to express the gravitational field in a cylindrical coordinate system with the symmetry axis of the spheroid as the $z$-axis and the plane perpendicular to the $z$-axis that contains the cloud centre as the $x$–$y$ plane (i.e. the $x$–$y$ plane coincides with the cloud's equatorial plane). Then, according to the elementary potential theory (Chandrasekhar 1942 and references therein), the gravitational potential inside the spheroid can be written as

$$\Phi(r,z) = \frac{3GM}{4a^3 e^2}\left(\frac{\arcsin e}{e} - \sqrt{1-e^2}\right) r^2$$
$$+ \frac{3GM}{2a^3 e^2}\left(\frac{1}{\sqrt{1-e^2}} - \frac{\arcsin e}{e}\right) z^2 - \frac{3GM}{2ae} \arcsin e, \quad (2)$$

where $M$, $a$ and $e$ denote the mass, the semimajor axis and the eccentricity of the spheroid. $G$ is the gravitational constant. The radial distance $r$ given in terms of the Cartesian coordinates, whose origin is the cloud's centre, is $r = \sqrt{x^2 + y^2}$. The mass of the spheroid depends on the density $\rho(t)$, as given by equation (1), and hence the potential of equation (2) depends on the time.

The components of the force due to the gravitational potential $\Phi(r, z)$ of the spheroid acting on a star located at $(r, z)$ are therefore $F_r = -\frac{\partial \Phi(r,z)}{\partial r}$ and $F_z = -\frac{\partial \Phi(r,z)}{\partial z}$. We consider that the $x$–$y$ plane of the spheroid is parallel or coincident with the Galactic plane and that from equation (2) the motion components in the $x$–$y$ plane can be obtained separately from the motion in the $z$ direction. For our purposes, it will suffice to study the star motions in the $x$–$y$ plane, and therefore, to use the force $F_r$ alone. We should add to $F_r$ a second term to take into account the action of the general gravitational field of the Galaxy. However, in our approximation, this term is neglected. Now we need to express the force $F_r$ in Cartesian coordinates. Using the elemental transformations $F_x = F_r\cos\theta$, $F_y = F_r\sin\theta$, where $\cos\theta = \frac{x}{r}$ and $\sin\theta = \frac{y}{r}$, and taking into account that the mass of the homogeneous spheroid is $M = \frac{4\pi \rho(t) a^3}{3}\sqrt{1-e^2}$, we obtain

$$F_x = -2\pi G \rho(t) \frac{\sqrt{1-e^2}}{e^2} \left(\frac{\arcsin e}{e} - \sqrt{1-e^2}\right) x \quad (3)$$

and

$$F_y = -2\pi G \rho(t) \frac{\sqrt{1-e^2}}{e^2} \left(\frac{\arcsin e}{e} - \sqrt{1-e^2}\right) y, \quad (4)$$

for $r \leq a$.

Substituting equation (1) into equations (3) and (4) and writing $k^2 = 2\pi G \rho_0 \frac{\sqrt{1-e^2}}{e^2}(\frac{\arcsin e}{e} - \sqrt{1-e^2})$, equations (3) and (4) reduce to $F_x = -k^2(1 + \lambda t)\, x$ and $F_y = -k^2(1 + \lambda t)\, y$.

The $x$–$y$ coordinate system is joined to the cloud and its origin is the centre of the cloud. We need to define now an inertial system of coordinates to which the movements of the cloud and stars can be referred. The $X$-axis of this second coordinate system agrees with the $x$-axis, except for the origin, and the $Y$-axis is parallel to the $y$-axis. We assume that the cloud is accelerated (or decelerated) as a whole only along the $X$-axis. The position of the cloud's centre in the $X$–$Y$ system is denoted by $(X_c(t), Y_c(t))$, where by definition $Y_c(t) = 0$. The transformations

$$X = x + X_c$$
$$Y = y \qquad (5)$$

convert the position $(x,y)$ of a star in the cloud to its position $(X,Y)$. The motion equations of a star lying in the cloud can be written $\ddot{X} = -k^2(1+\lambda t)(X-X_c)$, $\ddot{Y} = -k^2(1+\lambda t)Y$. Deriving twice relations (5) with respect to $t$, we have that $\ddot{X} = \ddot{x} + \ddot{X}_c$, $\ddot{Y} = \ddot{y}$. Then, and taking into account that $(X - X_c) = x$, the motion equations become

$$\ddot{x} + k^2(1 + \lambda t)x = \alpha \qquad (6)$$

$$\ddot{y} + k^2(1 + \lambda t)y = 0, \qquad (7)$$

where $\alpha = -\ddot{X}_c$, that is to say, an inertial force. For simplicity, we consider that $\alpha$ is a constant in each phase of the process. The motion equations (6) and (7) permit us to calculate orbits of stars, linked to the cloud, in the frame of reference with origin at the centre of gravity of the cloud. The solutions of equations (6) and (7) can be expressed in terms of the Airy functions (see Appendix A) as follows:

$$x = \pi(\dot{B}i(u_0) Ai(u) - \dot{A}i(u_0) Bi(u)) x_0$$
$$- \frac{\pi}{u_0 \lambda} (Bi(u_0) Ai(u) - Ai(u_0) Bi(u)) v_{0x}$$
$$- \frac{\pi \alpha}{(u_0 \lambda)^2} Ai(u) \int_{u_0}^{u_0(1+\lambda t)} Bi(u) du$$
$$+ \frac{\pi \alpha}{(u_0 \lambda)^2} Bi(u) \int_{u_0}^{u_0(1+\lambda t)} Ai(u) du \qquad (8)$$

$$y = \pi(\dot{B}i(u_0)Ai(u) - \dot{A}i(u_0)Bi(u)) y_0$$
$$- \frac{\pi}{u_0 \lambda}(Bi(u_0)Ai(u) - Ai(u_0)Bi(u)) v_{0y}, \qquad (9)$$

where $u = u_0(1 + \lambda t)$ and $u_0 = -(\frac{k}{\lambda})^{2/3}$. The dot over the Airy functions is here employed to denote the derivative with respect to the variable $u$. Deriving equations (8) and (9) with respect to $t$, we obtain

$$v_x = \pi x_0 u_0 \lambda (\dot{B}i(u_0)\dot{A}i(u) - \dot{A}i(u_0)\dot{B}i(u))$$
$$- \pi v_{0x}(Bi(u_0)\dot{A}i(u) - Ai(u_0)\dot{B}i(u))$$
$$- \frac{\pi \alpha}{u_0 \lambda} \dot{A}i(u) \int_{u_0}^{u_0(1+\lambda t)} Bi(u) du$$
$$+ \frac{\pi \alpha}{u_0 \lambda} \dot{B}i(u) \int_{u_0}^{u_0(1+\lambda t)} Ai(u) du, \qquad (10)$$

$$v_y = \pi y_0 u_0 \lambda(\dot{B}i(u_0)\dot{A}i(u) - \dot{A}i(u_0)\dot{B}i(u))$$
$$- \pi v_{0y}(Bi(u_0)\dot{A}i(u) - Ai(u_0)\dot{B}i(u)). \qquad (11)$$

Equations (8) and (9) give in parametric form the orbit of a star whose initial position and velocity are $(x_0, y_0)$ and $(v_{0x}, v_{0y})$, and equations (10) and (11) give the corresponding orbit velocities.

Since we will apply the former equations to calculate the orbits in phases A and C, we will distinguish the corresponding initial conditions, variables and parameters with superscript 'A' and 'C', respectively.

In the case of phase B, $\alpha = 0$ and $\lambda = 0$, and the solutions of the motion equations (6) and (7) are simply given by

$$x = x_0 \cos kt + \frac{v_{0x}}{k} \sin kt = A_x \sin(\beta_x + kt)$$
$$y = y_0 \cos kt + \frac{v_{0y}}{k} \sin kt = A_y \sin(\beta_y + kt), \qquad (12)$$

where $A_x = \sqrt{(x_0)^2 + (\frac{v_{0x}}{k})^2}$, $A_y = \sqrt{(y_0)^2 + (\frac{v_{0y}}{k})^2}$, $\beta_x = \arctan(\frac{kx_0}{v_{0x}})$ and $\beta_y = \arctan(\frac{ky_0}{v_{0y}})$. The velocity components are trivially derived from the equation system (12). When applying equations (12), we will denote their initial conditions, parameters and variables with the superscript 'B', referring to phase B.

### 2.2 Initial conditions: the velocity distribution of field stars in the protocloud, previous to the acceleration process

We assume that the field stars that are passing through the cloud at the beginning of phase A ($t = 0$) departed from distant regions of the Galactic disc where the gravitational influence of the protocloud was insignificant. There, the velocity distribution of the field stars is assumed to be Gaussian and written in the form

$$f_\infty(v_x, v_y, v_z) = \frac{N_\star}{(2\pi)^{3/2}\sigma_x\sigma_y\sigma_z} e^{-(\frac{v_x^2}{2\sigma_x^2} + \frac{v_y^2}{2\sigma_y^2} + \frac{v_z^2}{2\sigma_z^2})}, \qquad (13)$$

where $N_\star$ is the number of stars pc$^{-3}$, and $\sigma_x$, $\sigma_y$ and $\sigma_z$ are the velocity dispersions in the respective axes. Now, our problem is to find from the velocity distribution at large distances from the cloud, expressed by the symbol $\infty$ in equation (13), the velocity distribution of the field stars in the cloud. Since we will study only the distribution in the components $v_x$ and $v_y$, we integrate equation (13) with respect to $v_z$. With this aim, we define $\xi = \frac{1}{\sqrt{2\pi}\sigma_z} \int_{-v_{z\max}}^{v_{z\max}} e^{\frac{v_z^2}{2\sigma_z^2}} = \text{Erf}(\frac{v_{z\max}}{\sqrt{2}\sigma_z})$, which is an estimation of the fraction of stars that oscillate within the Galactic gas layer and hence can potentially penetrate the protocloud. For $v_{z\max} = 9$ km s$^{-1}$, the maximum height above and below the Galactic plane reached by stars is around 100 pc (see table 12-1 of Mihalas 1968), of the order of thickness of the gas layer of the Galactic disc, resulting $\xi \approx 0.35$. We have adopted $\sigma_z = 20$ km s$^{-1}$. Considering that $\sigma_x = \sigma_y = \sigma$, the distribution of equation (13) depends only on the magnitude of the velocity $\sqrt{v_x^2 + v_y^2}$, denoted here as $v_\infty$, which permits to express the velocity volume element $(dv_x dv_y dv_z)$ in cylindrical coordinates, facilitating the integration. Thus, equation (13) becomes

$$f_\infty(v_\infty) = \frac{\xi N_\star}{2\pi\sigma^2} e^{-\frac{v_\infty^2}{2\sigma^2}}. \qquad (14)$$

The next step is to obtain the velocity distribution of the field stars in a point $P_0$ of the protocloud. Let $P_1$ be a point close to $P_0$, lying both on the $(x, y)$ plane of the protocloud. If the segment $P_0P_1$ has a length $dl$, we define an area element $dS$ as $dS = dl dH$, where $dH$ is a height perpendicular to the $x$–$y$ plane. Besides, we assume that the area element (see Fig. 1) is crossed at the time $t = 0$ by stars moving in directions between $\phi$ and $\phi + d\phi$ with velocities in the range $v$ to $v + dv$. Therefore, the number of stars arriving at $P_0$ in the time interval d$t$, within the small range of directions (d$\phi$), velocities (d$v$) and time (d$t$) under consideration, is

$$dN = f(v)\, v\, d\phi\, dv\, dV\,,\qquad(15)$$

where $f(v)$ is the distribution function in the cloud, $v\,d\phi\,dv$ is the velocity volume element and $dV = dS\, v\cos\omega\, dt$ is the space volume element. $\omega$ is the angle between the velocity vector $\boldsymbol{v}$ and the segment d$l$. The magnitude of the velocity is naturally given by $v = \sqrt{v_x^2 + v_y^2}$.

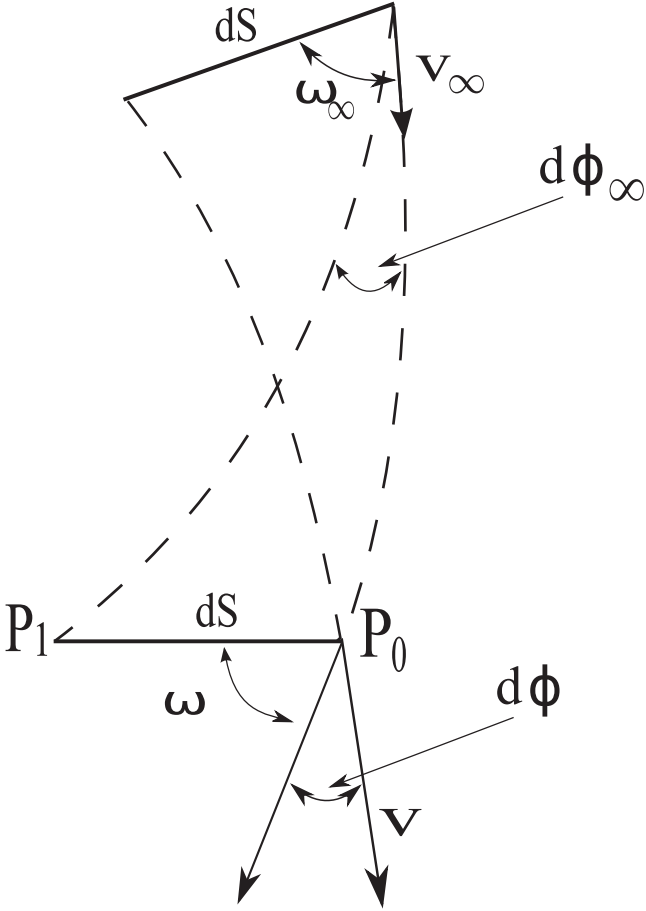

**Figure 1.** Schematic representation of orbits of field stars arriving to the cloud from infinity. The figure illustrates the fact that the number of stars in a given volume element in phase space is constant, which allows us to infer the field-star velocity distribution in the cloud from the field-star velocity distribution at infinity.

Tracing back in time representing orbits of the stars characterized by equation (15), we find that these stars crossed an area element $dS_\infty$ located at a large distance from the protocloud at a certain time (negative). Therefore, the following formula should be satisfied:

$$dN = f_\infty(v_\infty)\, v_\infty\, d\phi_\infty\, dv_\infty\, dV_\infty\,,\qquad(16)$$

where $f_\infty(v_\infty)$ is given by equation (14), $dV_\infty = dS_\infty\, v_\infty \cos\omega_\infty\, dt$ and the rest of the expressions has the same meaning as for the cloud's position $P_0$, but referred to a far position outside the protocloud (see Fig. 1).

If we impose the condition $dS_\infty = dS$, the calculation of the orbits, schematically represented in Fig. 1, shows that condition $\frac{d\phi_\infty}{d\phi}\frac{\cos\omega_\infty}{\cos\omega} = \frac{v}{v_\infty}$ should be fulfilled. For positions outside the spheroid, we should use the external potential (see for instance Mihalas 1968). From the law of conservation of energy, we have that

$v_\infty = \sqrt{v^2 + 2\Phi(r)}$, and hence $\frac{dv_\infty}{dv} = \frac{v}{v_\infty}$. Equalizing equations (15) and (16) and making the corresponding replacements, we obtain $f(v) = f_\infty(v_\infty)$. Note that the phase space volume is conserved, as stated by the Liouville's theorem. Thus,

$$f(v, r) = \frac{\xi N_\star}{2\pi\sigma^2} e^{-\frac{v^2 + 2\Phi(r)}{2\sigma^2}},\qquad(17)$$

for $v \geq \sqrt{2\,|\,\Phi(r)\,|} = v_{\rm esc}$, where $\Phi(r)$ is the potential given by equation (2) with $z = 0$. Note that the velocities of the field stars passing through the protocloud are greater that the escape velocity $v_{\rm esc}$ from the cloud. Since we have adopted $N_\star$ stars pc$^{-3}$ = constant in the Galactic disc, the number density of field stars within the cloud is too uniform. Indeed, $\int_0^{2\pi}\int_{v_{\rm esc}}^\infty v f(v,r) dv d\phi = \xi N_\star$. On the other hand, the velocity distribution depends on the position of the volume element in the cloud. Equation (17) will permit us to define the initial conditions of the field stars in the cloud at the beginning of the phase A in order to apply equations (8)–(11). The velocity dispersions of the field stars depend on the star ages and range between 20 and 40 km s$^{-1}$ (Nordström et al. 2004). From now on, we will use $\sigma = 35$ km s$^{-1}$, which corresponds approximately to the dispersion of the G-type stars that have ages of $\approx$5 Gyr. The star density in the solar neighbourhood is $\approx$0.08 stars pc$^{-3}$ (Allen 1963) and this number could be adopted for $N_\star$. However, we will use a much lower value for $N_\star(= 0.002$ stars pc$^{-3}$) to simplify the calculations and graphic representations of the data. Thus, the absolute values related to the star density can be simply estimated by multiplying our results by a factor $f_\star$ of $\approx$40.

### 2.3 Criterion of capture

Our rule to decide whether a star is captured by the cloud is that the star position and velocity at the end of phase A, and hence its initial conditions for phase B, determine that the star orbits during the phase B period within the physical limits of the cloud (i.e. $r \leq a$). This criterion excludes those stars that, even gravitationally captured (i.e. $v < v_{\rm esc}$), have orbits reaching the outer of the spheroid. However, this rather restrictive criterion simplifies greatly our calculations.

Let us now determine the possible trajectories of stars lying at the moment in which phase A starts in a volume element of the cloud located at $(x_0^A, y_0^A)$. Hence, these stars have in common the same initial position, but different initial velocities. Denoting the magnitude of the initial velocity of a star by $v$ and the angle between the $x$-axis and the velocity by $\phi$, the corresponding Cartesian components can be written

$$vx_0^A = v\cos\phi$$
$$vy_0^A = v\sin\phi\,,\qquad(18)$$

where $\phi$ is measured anticlockwise from the positive $x$-axis. By varying $v$ and $\phi$ we cover all possible initial velocities in accordance with the velocity distribution (17) of the volume element in consideration. Denoting the duration of phase A by $t_f^A$ and substituting the position initial $(x_0^A, y_0^A)$ with a pair of numerical values, equations (18), and $t = t_f^A$ in equations (8)–(11), we obtain the position and velocity of a star at the end of phase A as a function of $v$ and $\phi$. That is to say $x_f^A(v, \phi), y_f^A(v, \phi), v_{fx}^A(v, \phi)$ and $v_{fy}^A(v, \phi)$), where the superscript refers to phase A and the subscript 'f' to the final time of the phase. Making $x_0^B = x_f^A(v, \phi), y_0^B = y_f^A(v, \phi), v_{0x}^B = v_{fx}^A(v, \phi)$ and $v_{0y}^B = v_{fy}^A(v, \phi)$ and replacing into equations (12), we have the possible orbits during phase B as a function of $(v, \phi)$ of the stars contained in the original volume element $(x_0^A, y_0^A)$ of phase A. Adopting the same criterion as above for the notation in order to distinguish the magnitudes of the different phases, the radial distance of an orbit can be written $r = \sqrt{(x^B)^2 + (y^B)^2}$, which is a function of $v$, $\phi$ and $t^B$. A good approximation for the maximum radial distance that can be reached during phase B is given by $r_{\max} \approx \sqrt{A_x^2 + A_y^2}$ (see equations 12). Since our capture criterion requires that $r_{\max} \leq a$, the equation $r_{\max} = a$ permits us to obtain the maximum velocity $v_{\max}$ that fulfils the capture criterion as a function of $\phi$. Choosing a fixed set of the model parameters ($a$, $e$, $\rho_0^A$, $\lambda^A$, $\alpha^A$, $t_f^A$), we can plot $v_{\max} = f(\phi)$ for each volume element ($x_0^A$, $y_0^A$) of the cloud. This means that a star moving in a given direction $\phi$ is captured if its initial velocity $v$ satisfies the condition $v_{\text{esc}} \leq v \leq v_{\max}$, where $v_{\text{esc}}$ is the escape velocity from the position ($x_0^A$, $y_0^A$) of the volume element under consideration (see Figs 2a and b). Therefore, the percentage of captured stars can be calculated from

$$P(r) = \frac{1}{\xi N_\star} \int_{\phi_{\min}(r)}^{\phi_{\max}(r)} \int_{v_{\text{esc}}(r)}^{v_{\max}(\phi,r)} v f(v, r) \mathrm{d}v \mathrm{d}\phi, \quad (19)$$

where $f(v, r)$ is given by equation (17). Here we write the initial position ($x_0^A$, $y_0^A$) as $r$ for short. To express $P(r)$ in per cent, equation (19) should be multiplied by 100. Figs 2(a) and (b) show, for two positions within the cloud, the corresponding curves $v_{\max}(\phi)$ in their respective ranges of $\phi_{\min} \leq \phi \leq \phi_{\max}$ in which the condition for capture is satisfied. In the rest of the $\phi$ domain, $v_{\max}(\phi)$ is imaginary or has values lower than $v_{\text{esc}}$. Since the maximum of $v_{\max}(\phi)$ lies on or close to $\phi = 0$, the angles of the fourth quadrant are considered negative (i.e. $= \phi - 360°$). If the cloud is decelerated, the maximum of the function occurs on the opposite direction, $\phi \approx 180°$. The partial integration of $f(v, r)$ between any two consecutive vertical lines, indicated by broken lines in Fig. 2, is equal to 1.

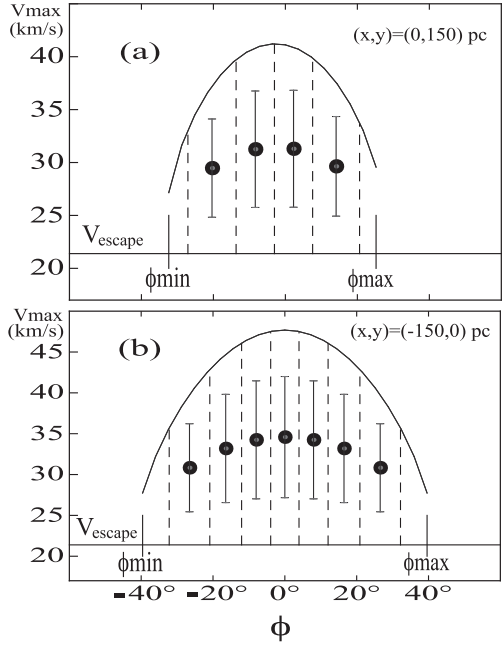

**Figure 2.** Diagrams showing the velocity directions ($\phi$) and velocity magnitudes ($v$) of the stars that are located at the initial position ($x$, $y$) on the cloud and that will be captured. The escape velocity from the position ($x$, $y$) on the cloud, with the cloud at rest, is indicated. The field stars within the space $\phi$–$v$ enclosed by the curve $v_{\max}(\phi)$ and the line indicating the escape velocity will be captured. The points and error bars show the mean velocities and velocity dispersions of the stars to be captured. The upper diagram corresponds to the position ($x$, $y$) = (0, 150) pc and the lower one to the position ($x$, $y$) = (− 150, 0) pc.

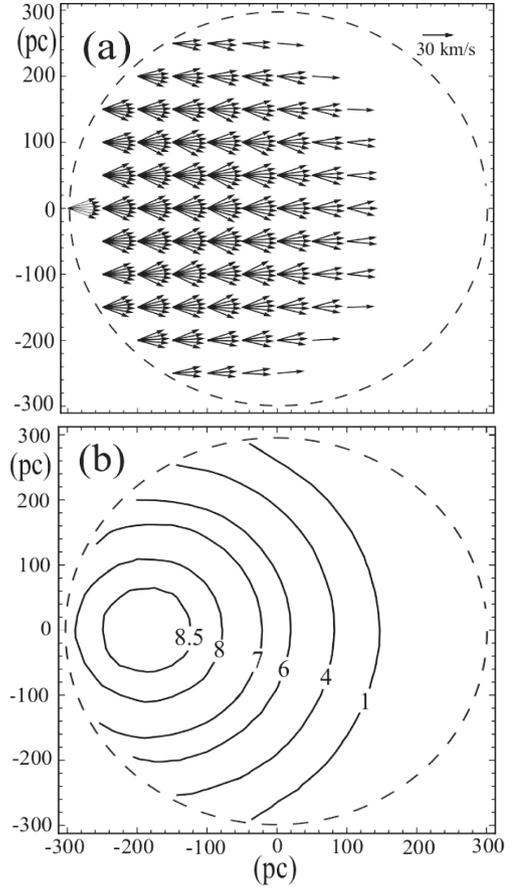

**Figure 3.** Panel (a) shows the initial mean velocities of the stars that will be captured for each volume element of 50 pc × 50 pc × 50 pc of the cloud's equatorial plane. The broken lines indicate the edge of the cloud. The velocity scale is given by the arrow (=30 km s$^{-1}$) near the top-right corner. Panel (b) shows the initial spatial distribution of percentages of the stars that will be captured at the end of the first phase (phase A). The broken line indicates the edge of the cloud.

Therefore, in the phase space volume represented by each slab there is a star whose probable velocity is represented by the mean velocity and velocity dispersion of the corresponding slab (see Figs 2a and b), where the mean velocities and dispersions are indicated. In the examples of Figs 2(a) and (b), there are four stars ($\approx \xi N_\star * \mathrm{d}V * P(r)$) and seven stars, respectively, that can be captured. For the volume element of the examples given in Fig. 2, we have taken $\mathrm{d}V = 50$ pc × 50 pc × 50 pc. In order to obtain the distribution of initial velocities and positions, on the whole cloud, of the finally captured stars, we divide the inner equatorial plane of the cloud into cells of area $\Delta x \times \Delta y = 50$ pc × 50 pc and adopt, for each cell (or volume element), a height of 50 pc, perpendicular to the $x$–$y$ plane. Fig. 3(a) displays the initial mean velocities in the corresponding central directions ($\phi_c$) of the captured stars, obtained by means of the above explained method. Here, we have taken the central position of each cell as the initial positions of the stars contained in the cell. Fig. 3(a) shows that the captured stars have initial velocities in directions close to the positive $x$-axis, direction in which the cloud is accelerated. From equation (19), we calculate the distribution of the percentage of capture as a function of the initial positions of the finally captured stars (Fig. 3b). The stars located, at the beginning of phase A, at the rear of the cloud are captured in greater numbers. If the cloud is decelerated in phase A, the just mentioned distributions are given by the mirror images of Figs 3(a) and (b). We have used the following set of values for the model parameters, namely ($a$, $e$, $\rho_0^A$, $\lambda^A$, $\alpha^A$, $t_f^A$) = (300 pc, 0.866, 7 atoms cm$^{-3}$, $\frac{1}{10}$ Myr$^{-1}$, $-3.0$ km s$^{-1}$ Myr$^{-1}$, 10 Myr). In the next section, we will give some justifications on the use of these particular values for the parameters of the model.

## 3 RESULTS

This model certainly permits the use of different values for the main parameters, as for instance the cloud dimension given by the semi-axis $a$ and the eccentricity $e$ or the cloud acceleration contained in the parameter $\alpha$. In the next subsection, we will study the effects of varying the parameters within their physical limits. However, we require a priori that the cloud volume, $\Omega = \frac{4}{3}\pi a^2 b$, be equal to the volume of a cylinder given by $\pi a^2 H$, where $b$ is the semiminor axis of the spheroid and $H$ the thickness of the H I Galactic layer ($\approx 200$ pc). Hence, $b = \frac{3}{4}H$. Since $b = \sqrt{1-e^2}$, $e = \sqrt{1-(\frac{3H}{4a})^2}$. Then, the value of $e$ depends on the value adopted for $a$. On the other hand, we consider that the protocloud is dynamically stable and in consequence has a minimum density $\rho_0 = \rho^\star$, where the so-called critical density $\rho^\star = \frac{0.165}{\beta_1}(M_\odot \text{ pc}^{-3}) = \frac{6.6}{\beta_1}$(atoms cm$^{-3}$) and $\beta_1$ is a pure number that depends on $\frac{b}{a} = \frac{3H}{4a}$ (see table 12$^\star$ of Chandrasekhar 1942).

We think of the interstellar cloud of this model as the progenitor of a gas–star complex. Elmegreen & Elmegreen (1983) found interstellar clouds with dimensions as large as 1 kpc and suggested that these superclouds may originate giant star complexes. The gas–star complexes have typical sizes of $\approx 650$ pc (Alfaro, Cabrera-Cano & Delgado 1992a,b; Efremov 1995, 2010). In accordance with these facts, we adopt $a = 300$ pc and in consequence $e = 0.866$ and $\rho_0^A = \rho^\star = 7$ atoms cm$^{-3}$.

### 3.1 Mean percentage of capture as a function of the parameters of the model

The distribution of the percentages of capture is not uniform over the cloud. This has a maximum at the back side of the cloud. In the example of Fig. 3(b), the maximum percentage of capture is $\approx 9$ per cent, while the mean percentage is $\approx 4$ per cent, averaged over the whole cloud. In other words, $\overline{P} = \frac{\int_\Omega P(r)dV}{\int_\Omega dV}$, where $P(r)$ is given by equation (19). In this subsection, we characterize each chosen set of parameters of the model by its resulting mean percentage of capture.

We denote the parameter $\alpha$, which corresponds to the cloud acceleration (with the sign minus), by $\alpha^A$ during the phase A. Then, we can write $\alpha^A$ in terms of the velocity of the cloud $V_f^A$, reached at the end of phase A ($t = t_f^A$) as $\alpha^A = -\frac{V_f^A - V_0^A}{t_f^A}$, where the initial velocity $V_0^A = 0$. From equation (1), the mass accretion rate of phase A, $\lambda^A$, can be expressed in terms of the final density $\rho_f^A$ of this phase as $\lambda^A = \frac{\rho_f^A - \rho_0^A}{\rho_0^A} \frac{1}{t_f^A}$. That is to say that the set of parameters of phase A, namely ($\alpha^A$, $\lambda^A$, $t_f^A$) is equivalent to the set ($V_f^A$, $\rho_f^A$, $t_f^A$). Setting $V_f^A$ to a certain value, we can represent the mean percentage of capture as a function of $t_f^A$, $\overline{P}(t_f^A)$, for different values of $\rho_f^A$ (see Fig. 4).

Similarly, fixing $\rho_f^A$ to a certain value, we can represent $\overline{P}(V_f^A)$, for different values of $t_f^A$ (see Fig. 5).

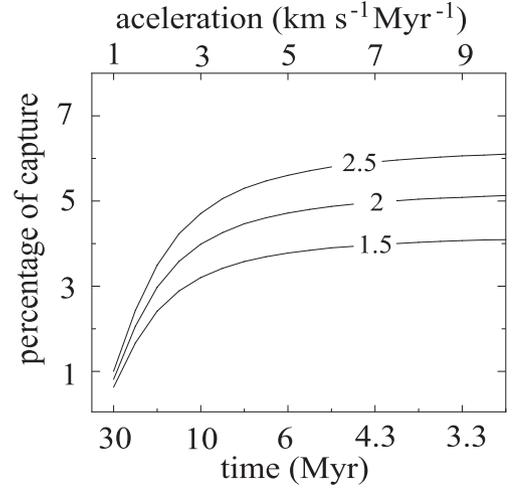

**Figure 4.** The mean percentage of captured field stars as a function of the duration of phase A ($t_f^A$) for the final densities of the cloud at the end of this phase $\rho_f^A = 1.5$, 2 and 2.5 times the initial density $\rho_0^A(= \rho^\star = 7$ atoms cm$^{-3}$). The final velocity of the cloud at the end of this phase is fixed at $V_f^A = 30$ km s$^{-1}$, velocity gained by the cloud during phase A due to a constant acceleration. The upper abscissa shows the corresponding accelerations.

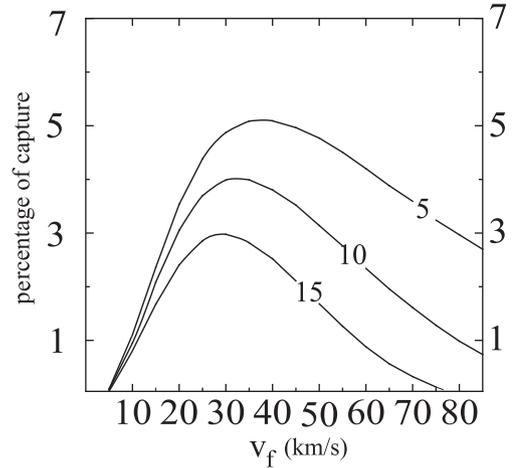

**Figure 5.** The mean percentage of captured field stars as a function of the final velocity $V_f^A$ of the cloud at the end of phase A for different durations of phase A: $t_f^A = 5$, 10 and 15 Myr. The final densities of the cloud at the end of this phase is fixed at $\rho_f^A = 2\rho^\star$ (i.e. two times the initial density $\rho_0^A$), where $\rho^\star = 7$ atoms cm$^{-3}$.

The curves displayed by Fig. 4 show that the lower the time of acceleration the greater the percentage of capture. Also an increase of $\rho_f^A$ implies an increase of $\overline{P}(t_f^A)$. We adopt $\rho_f^A = 2\rho_0^A = 2\rho^\star$, as a moderate value. Thus, the total mass of the cloud at the end of phase A is $\approx 2 \times 10^7$ M$_\odot$, which is within the range of the supercloud masses (Elmegreen & Elmegreen 1983). Besides, note that these curves have a turnover point at $t_f^A \approx 10$ Myr, from which the slope chances abruptly. Fig. 5 shows that the maximum of the percentage of capture occurs at $V_f^A \approx 30$–$40$ km s$^{-1}$, which agrees approximately with the velocity dispersion of field stars, $\sigma$.

We will adopt $t_f^A = 10$ Myr, although we have not a strong observational or physical justification, except that this time coincides roughly with the time needed for interstellar clouds to cross a spiral arm. We will adopt $V_f^A = 30\,\mathrm{km\,s^{-1}}$, which is compatible with the velocity change experienced by gas clouds during the passage through an spiral arm (Roberts 1969, 1970). Hence, we adopt from now on $(V_f^A, t_f^A, \rho_f^A) = (30\,\mathrm{km\,s^{-1}}, 10\,\mathrm{Myr}, 2\rho^\star)$.

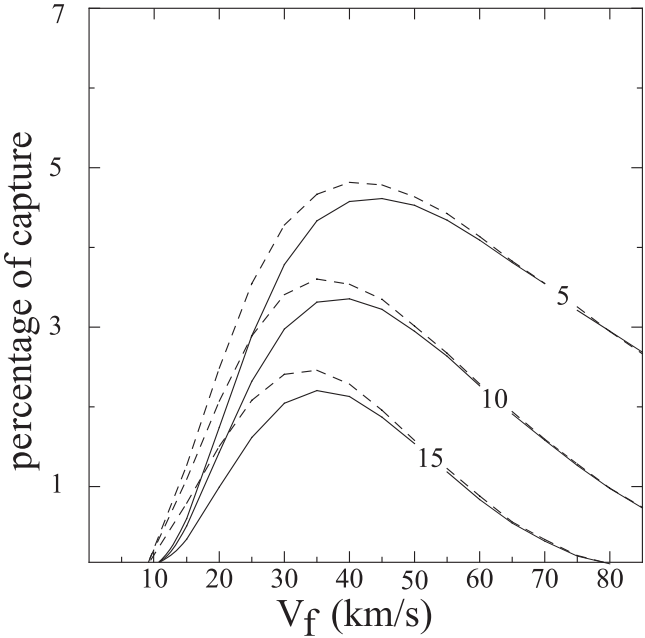

**Figure 6.** The same as Fig. 5, but calculated under the interpretation that the cloud's density increment is due to a process of compression and gravitational contraction. The broken-line curves represent the results obtained with the cloud initial density $= \frac{1}{2}\rho^\star$, and the full-line curves those obtained with the cloud initial density $= \rho^\star$. In both cases, the final density is the same and equals $2\rho^\star$.

We have above assumed that the cloud's volume is constant. This consequently implies that the increment of the cloud's density should be entirely due to mass accretion, process in which the density would increase exponentially with time. However, we have adopted a linear law of density increment (equation 1). In the appendix, we demonstrate that the use of an exponential law for the cloud density variation, $\rho(t) = \rho_0 e^{\mu t}$ with $\mu = $ constant, yields results that are almost coincident with those we have obtained using equation (1).

Our model can too consider the case in which the cloud's density increment during phase A is purely due to a homologous contraction of the cloud. We assume that the cloud contracts from the initial semimajor axis $a_0$ to the final one $a_f$. We here analyse the capture of the field stars lying at $t = 0$ within the cloud's inner volume delimited by the surface of an ellipsoid of semimajor axis $a_f$ (i.e. within the final volume that the cloud reaches at the end of the contraction phase). The escape velocity $v_{esc}$, used to determine the initial velocity distribution of the star field, must be calculated with $a = a_0$ and the capture criterion with $a = a_f$. Since the cloud's mass $M = \frac{4\pi\rho(t)a^3}{3}\sqrt{1-e^2}$ is here constant and if the density distribution remains homogeneous during the cloud's contraction, $a_0 = a_f(1 + \lambda^A t_f^A)^{\frac{1}{3}} = (\frac{\rho_f^A}{\rho_0^A})^{\frac{1}{3}} a_f$. If $a_f = 300\,\mathrm{pc}$ as adopted above for the size of the stable cloud and $\frac{\rho_f^A}{\rho_0^A} = 2$, we obtain $a_0 = 380\,\mathrm{pc}$ and the corresponding capture percentages shown in Fig. 6 by full lines. We have adopted the critical density $\rho^\star$ as the initial density of the cloud (i.e. $\rho_0^A = \rho^\star$). However, the model admits that the initial density can be lower. The model only requires that the density of the cloud at the end of phase A ($\rho_f^A$) must be greater than the critical density $\rho^\star$. For instance, if $\rho_0^A = \frac{1}{2}\rho^\star$ and $\rho_f^A = 4\rho_0^A = 2\rho^\star$, we obtain $a_0 = 480\,\mathrm{pc}$ and the corresponding capture percentages shown in Fig. 6 by broken lines. Note that these results of Fig. 6 are only slightly lower than those of Fig. 5.

Detailed studies of the formation and evolution of an interstellar gas cloud requires the realization of complex hydrodynamical calculations with the inclusion of gravitational and magnetic forces, shock waves, heating and cooling processes, turbulent motions, the cloud's rotation etc., which is out of the scope of this work. Nevertheless, we have shown above that the overall results on the capture of field stars are rather insensitive to the evolutionary details of the cloud.

### 3.2 Spatial distribution of positions and velocities of the captured field stars at the beginning and end of phases A and B

In this and the following sections, we consider that $a$ and $e$ remain constant during the three phases. Table 1 summarizes the values of the different parameters used in the simulations of each phase. In Section 2.3, we have defined the conditions of capture. For each considered volume element d$V$ of the cloud, we make calculations similar to those represented in Figs 2(a) and (b). These figures delimit each initial velocity space that permits the capture of a star, and give the corresponding mean velocity $\bar{v}$ and dispersion $\sigma^\star$ that characterize the probable initial velocity of the star. In the following simulations, the initial velocity $v$ of the captured star and the angle $\phi$ (equation 18) are given by a number at random between $\bar{v} - \sigma^\star$ and $\bar{v} + \sigma^\star$ and between $\phi_c - \frac{d\phi}{2}$ and $\phi_c + \frac{d\phi}{2}$, respectively. d$\phi$ is the width of the corresponding vertical strip between the lines $\phi = \phi_c - \frac{d\phi}{2}$ and $\phi = \phi_c + \frac{d\phi}{2}$, indicated by broken lines in the examples of Figs 2(a) and (b). For the position $x_0^A$, $y_0^A$, we choose a number at random between $x_0^A - \frac{\sqrt[3]{dV}}{2}$ and $x_0^A + \frac{\sqrt[3]{dV}}{2}$, and between $y_0^A - \frac{\sqrt[3]{dV}}{2}$ and $y_0^A + \frac{\sqrt[3]{dV}}{2}$. Fig. 7(a) displays the initial positions and velocities of the star to be captured.

Remembering that $\alpha^A = -\frac{V_f^A - V_0^A}{t_f^A}$ and $\lambda^A = \frac{\rho_f^A - \rho_0^A}{\rho_0^A}\frac{1}{t_f^A}$, and that we have adopted $(V_f^A, t_f^A, \rho_f^A) = (30\,\mathrm{km\,s^{-1}}, 10\,\mathrm{Myr}, 2\rho_0^A)$, we have the numerical values for all parameters and initial conditions needed to apply equations (8)–(11). Thus, we obtain the positions and velocities of the captured field stars at the end of phase A (Fig. 7b). The distributions represented by Figs 7(a) and (b) correspond to stars captured within the cloud's equatorial disc of thickness $\Delta H (= 50\,\mathrm{pc})$. In our model, the spatial and velocity distributions of the captured stars in the cloud's layers of different heights $z$ are all equal. Hence, to obtain the total number of stars captured by the cloud, this relative number of captured stars (=362) should be multiplied by $f_\star \frac{H}{\Delta H} \approx 160$.

In order to calculate the star orbits in phase B by means of equations (12), we should take into account the continuity conditions prescribing that $x_0^B = x_f^A$, $y_0^B = y_f^A$, $v_{0x}^B = v_{fx}^A$, $v_{0y}^B = v_{fy}^A$, $\rho_0^B = \rho_f^A$. Since the orbits are periodic in phase B, it not necessary to define the absolute duration of phase B; only the fraction of the orbit period $T (= \frac{2\pi}{k^B} \approx 94\,\mathrm{Myr})$ in which the phase finishes. We give four possibilities for the relative durations of the phase, namely $t_f^B(i) = i\frac{T}{4} + nT$, where $i = 0, 1, 2$ and 3, and $n = 0, 1, 2, \ldots$ We leave $n$ undefined. The position and velocity distribution corresponding to $t_f^B(0)$ agrees with the one at the end of phase A (Fig. 7b).

**Table 1.** Synopsis of the main parameters of the model. The semimajor axis *a* and the eccentricity *e* of the cloud are considered constant in the three phases and equal to 300 pc and 0.866, respectively.

| Parameter | Units | Phase A symbol | Value | Phase B symbol | Value | Phase C symbol | Value case (1) | Value case (2) |
|---|---|---|---|---|---|---|---|---|
| Duration of the phase | Myr | $t_f^A$ | 10 | $t_f^B$ | undetermined | $t_f^C$ | 10 | 10 |
| Cloud's velocity at the phase's start | km s$^{-1}$ | $V_0^A$ | 0 | $V_0^B$ | 30 | $V_0^C$ | 30 | 30 |
| Cloud's velocity at the phase's end | km s$^{-1}$ | $V_f^A$ | 30 | $V_f^B$ | 30 | $V_f^C$ | 0 | 30 |
| Acceleration of the cloud | km s$^{-1}$ Myr$^{-1}$ | $-\alpha^A$ | $3.0 = \frac{V_f^A - V_0^A}{t_f^A}$ | $-\alpha^B$ | 0 | $-\alpha^C$ | $-3.0 = \frac{V_f^C - V_0^C}{t_f^C}$ | 0 |
| Cloud's density at the phase's start | atoms cm$^{-3}$ | $\rho_0^A$ | 7 | $\rho_0^B$ | 14 | $\rho_0^C$ | 14 | 14 |
| Cloud's density at the phase's end | atoms cm$^{-3}$ | $\rho_f^A$ | 14 | $\rho_f^B$ | 14 | $\rho_f^C$ | 0 | 0 |
| Growth rate of the cloud's density | Myr$^{-1}$ | $\lambda^A$ | $0.1 = \frac{\rho_f^A - \rho_0^A}{\rho_0^A}\frac{1}{t_f^A}$ | $\lambda^B$ | 0 | $\lambda^C$ | $-0.1 = \frac{\rho_f^C - \rho_0^C}{\rho_0^C}\frac{1}{t_f^C}$ | $-0.1$ |

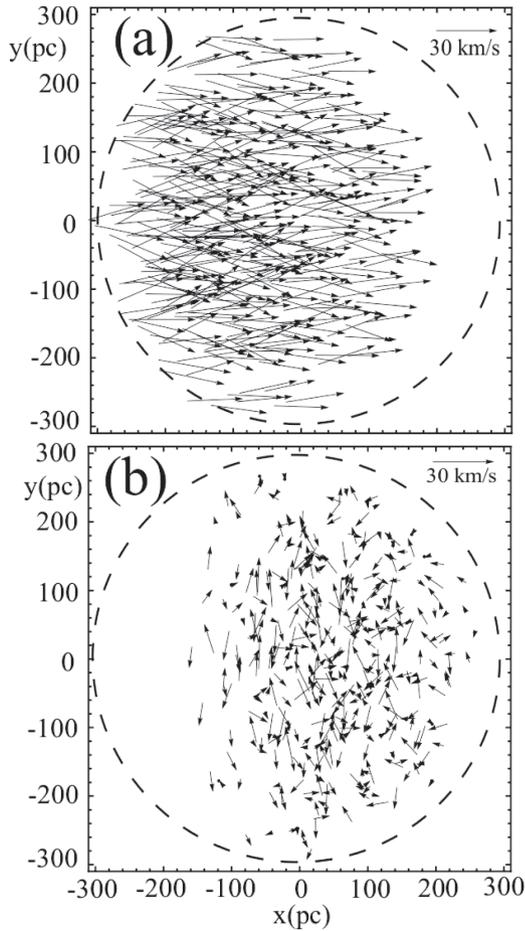

**Figure 7.** Panel (a): spatial distribution and velocity field at the start of the first phase (phase A) for the field stars that will be captured at the end of phase A. The arrows trace the velocity field of these stars. The velocity scale is given by the arrow (=30 km s$^{-1}$) near the top-right corner. The broken lines indicate the edge of the cloud. Panel (b): spatial distribution and velocity field of the captured stars at the end of phase A.

### 3.3 Space distribution of positions and velocities of the captured field stars at the end of phase C

A summary of the parameter values used in this section is given in last columns of Table 1. Since the density of the cloud does not change during phase B, $\rho_0^C = \rho_f^A$. Using also the final star positions and velocities of phase B as the initial conditions of phase C, we can apply equations (8)–(11) to calculate the star orbits in the phase under consideration. We take conveniently $t_f^C = 10$ Myr, for the duration of phase C; with which the star orbits remain within the cloud borders where our equations are valid. We consider two possibilities for the development of phase C that lead to similar results: (1) the cloud is decelerated (or accelerated, if it has been decelerated during phase A) and loses all its mass; (2) the cloud loses all its mass, but here the velocity of the cloud as a whole is unaffected. Therefore for case (1), we have that $\alpha^C = -\frac{V_f^C - V_0^C}{t_f^C}$ and $\lambda^C = \frac{\rho_f^C - \rho_0^C}{\rho_0^C}\frac{1}{t_f^C}$, where $\rho_f^C = 0$ and $V_0^C = V_f^A$. If $V_f^C = 0$, the cloud in its last stage of disintegration is at rest in the adopted inertial system of coordinates. Hence, the star velocities given by equations (10) and (11) at the end of phase C coincide with the star velocities referred to the inertial system, which could be assimilated to the LSR. Then $\alpha^C = -\frac{V_f^A}{t_f^C}$, $\lambda^C = -\frac{1}{t_f^C}$ and $u^C = 0$. Substituting $u = u^C = 0$ and $u_0 = u_0^C$ into the last two terms of equations (8) and (10), we find that $-\frac{\pi\alpha}{(u_0\lambda)^2}Ai(u)\int_{u_0}^{u}Bi(u)du + \frac{\pi\alpha}{(u_0\lambda)^2}Bi(u)\int_{u_0}^{u}Ai(u)du = \frac{1}{2}\frac{\alpha^C}{(\lambda^C)^2} = \frac{1}{2}V_f^A t_f^C$ and that $-\frac{\pi\alpha}{u_0\lambda}\dot{Ai}(u)\int_{u_0}^{u}Bi(u)du + \frac{\pi\alpha}{u_0\lambda}\dot{Bi}(u)\int_{u_0}^{u}Ai(u)du = -\frac{\alpha^C}{\lambda^C} = V_f^A$. Thus, the $x$-component of the position and of the velocity in case (1) can be written $x_f^C(1) = \chi(1) + \frac{1}{2}V_f^A t_f^C$ and $v_{fx}^C(1) = v(1) + V_f^A$, where $\chi(1)$ and $v(1)$ represent the first two terms of equations (8) and (10), respectively. Since $\alpha^C = 0$ in case (2), the two last terms of equations (8) and (10) are null and therefore $x_f^C(2) = \chi(2)$ and $v_{fx}^C(2) = v(2)$. In order to refer the velocity $v_{fx}^C(2)$ to the inertial system, the velocity $V_f^C = V_f^A$ should be added to $v_{fx}^C(2)$. For a given star $\chi(1) = \chi(2)$ and $v(1) = v(2)$, hence the final velocity in $x$ agrees in both cases and the $x$-component of the position differs only in a constant. On the other hand, $y_f^C(1) = y_f^C(2)$ and $v_{fy}^C(1) = v_{fy}^C(2)$. In other words, cases (1) and (2) are equivalent. The distribution of positions and velocities of the captured stars at the final of phase C depends on the adopted $t_f^B(i)$ (see Section 3.2). Figs 8(a) and (b) show the final star positions and velocities of phase C derived with $t_f^B(0)$ and with $t_f^B(2)$, respectively. The velocity distribution corresponding to Fig. 8(a) (with plus symbols) and to that of Fig. 8(b) (with point symbols) are shown in Fig. 9.

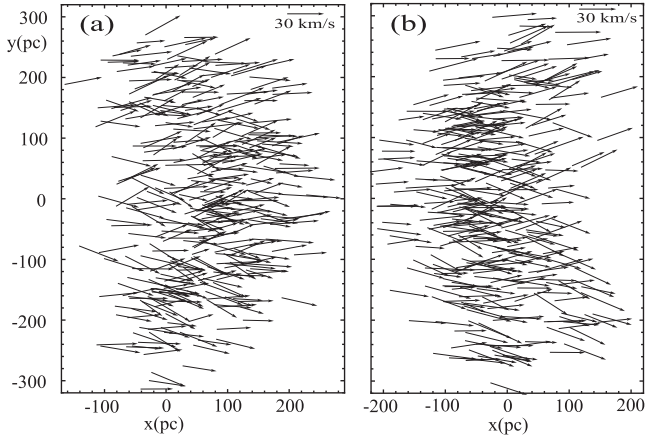

**Figure 8.** Spatial distributions and velocity fields for the captured field stars at the end of the final phase (phase C). The positions, magnitudes and directions of the vectors indicate the positions and velocities of the stars. We here draw a representative fraction of the total number of captured stars, which multiplied by a factor of ≈160 gives the total number of captured stars. The distributions are obtained with the duration of phase B given by $t_f^B(0)$ (panel a), and by $t_f^B(2)$ (panel b).

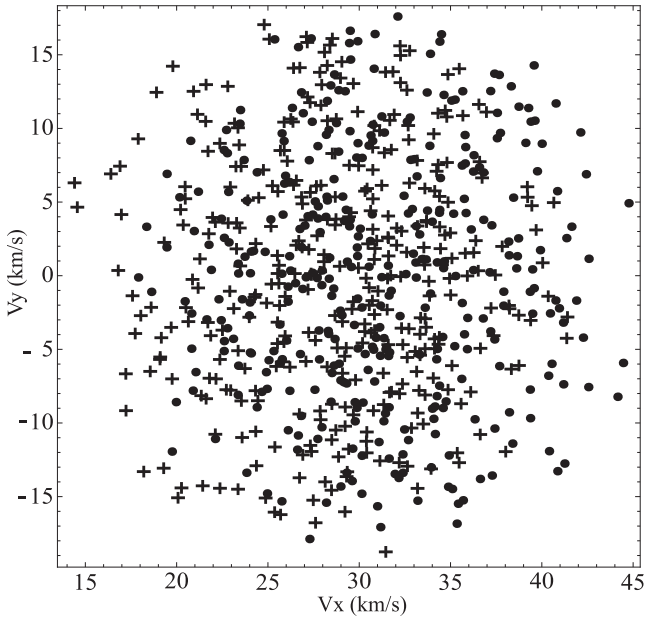

**Figure 9.** Distribution of velocity components corresponding to the velocity field of Fig. 8(a) (plus symbol) and of Fig. 8(b) (point symbol).

Figs 10(a) and (b) show the final star positions and velocities of phase C derived with $t_f^B(1)$ and with $t_f^B(3)$, respectively. The velocity distribution corresponding to Fig. 10(a) (with plus symbols) and to that of Fig. 10(b) (with point symbols) are shown in Fig. 11.

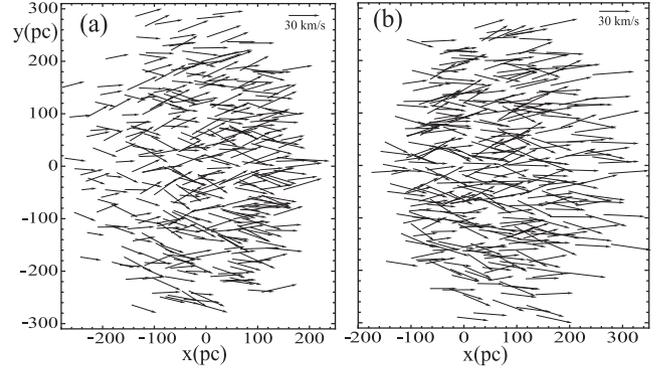

**Figure 10.** Spatial distributions and velocity fields for the captured field stars at the end of the final phase (phase C). The positions, magnitudes and directions of the vectors indicate the positions and velocities of the stars. We here draw a representative fraction of the total number of captured stars, which multiplied by a factor of ≈160 gives the total number of captured stars. The distributions are obtained with the duration of phase B given by $t_f^B(1)$ (panel a), and by $t_f^B(3)$ (panel b).

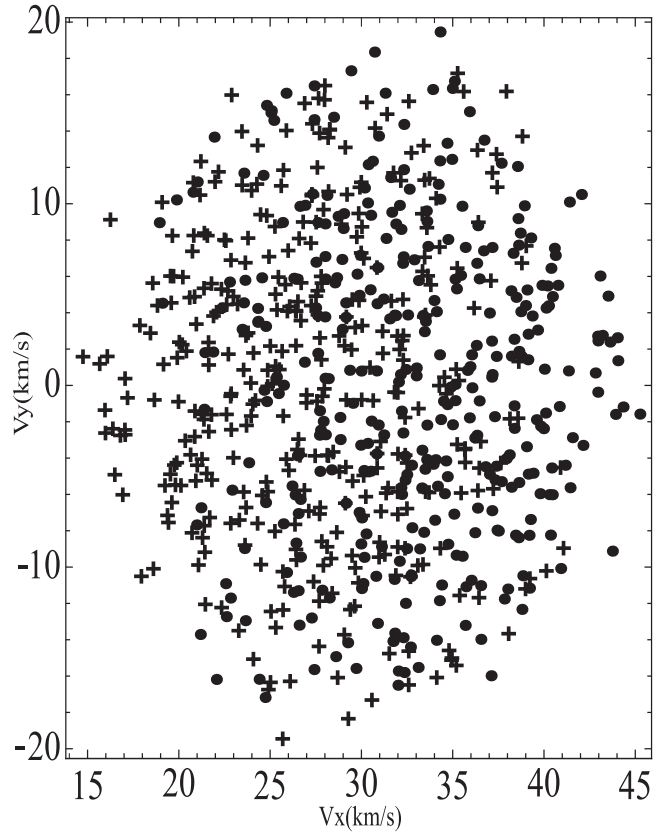

**Figure 11.** Distribution of velocity components corresponding to the velocity field of Fig. 10(a) (plus symbol) and of Fig. 10(b) (point symbol).

### 3.4 The capture of the Sun

A question that we should answer is whether the Sun's position among the local moving groups is by chance or not. There is a fact that might indicate that it is not by chance. The Sun's velocity with respect to the LSR is much lower than the velocity dispersion of the G-type stars. This could indicate that the Sun has interacted with an interstellar cloud. This possibility is of great interest because of its possible effects on the Earth's environment (Clube & Napier 1986; Bailey, Clube & Napier 1990).

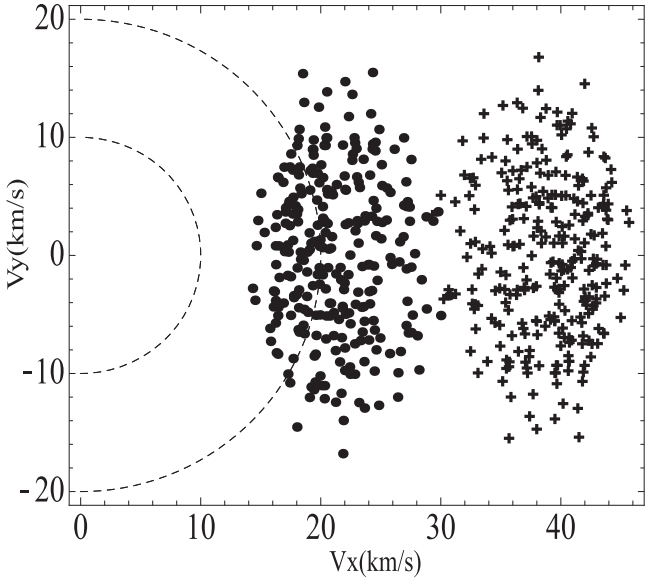

**Figure 12.** Distribution of velocity components for the captured field stars of solar type at the end of the final phase. The plus symbols correspond to the case $t_f^B(0)$ and the point symbols to the case $t_f^B(2)$. The region between the two concentric circles traced by broken lines corresponds to the velocity range in which the LSR solar velocity lies.

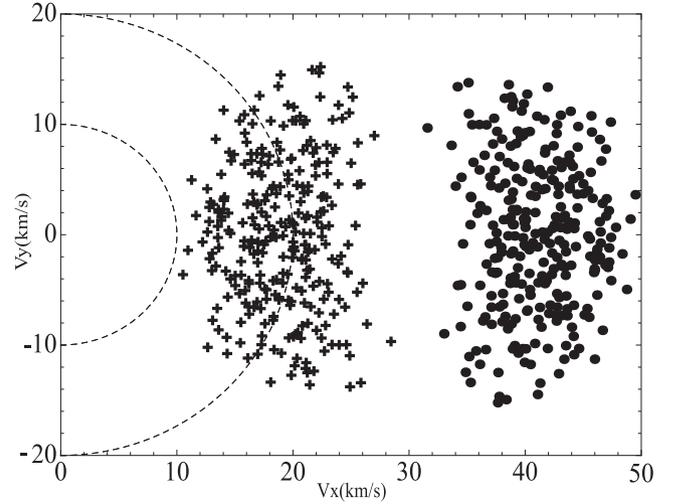

**Figure 13.** The same as Fig. 12 but for the case $t_f^B(1)$ (plus symbols) and for the case $t_f^B(3)$ (point symbols).

So the relatively low LSR-velocity of the Sun could be explained by means of a mechanism of capture. The Sun met an interstellar cloud in coincidence with the passage of a spiral arm that decelerated suddenly the cloud. Another possibility is that the collision of an HVC with the Galactic disc accelerated a concentration of Galactic gas, through which the Sun was circumstantially passing. In the course of the last 400 Myr, a flow of HVCs coming from the Magellanic Clouds might have collided with the gas layer of the Galactic disc (Olano 2004). We propose that the same cloud that captured the Sun captured the field stars of the local moving groups.

Figs (8)–(11) represent the distributions of positions and velocities for the captured field stars of all spectral types, at the end of phase C. Here our aim is to analyse these distributions only for the captured field stars of solar type (i.e. G-type stars). Since the field stars of solar type have ages of 5 Gyr and velocity dispersions $\geq 35$ km s$^{-1}$ (Nordström et al. 2004), following the method of Section 2.3, we use the lower limit of integration given by $v_G = \sqrt{v_\infty^2 - 2\Phi(r)}$, where $v_\infty = 35$ km s$^{-1}$. Note that when $v_\infty = 0$, $v_G = v_{esc}$ (see equation 19), and that the $\phi$ range of integration depends too on the value of $v_G$ (see Fig. 2). In order to derive the distribution of positions and velocities for this particular type of stars, we use $t_f^B(i)$ of phase B, where $i=0, 1, 2$ and 3, and case (2) of phase C. For the rest of the parameters, we use the same values as before, except for $N_\star$, where we use $N_\star = 0.02$ stars pc$^{-3}$ (i.e. we should multiply by a factor $f_\star = 4$ to obtain the star density observed in the solar vicinity). The resulting distributions are displayed in Figs 12 and 13. Since the LSR velocity of the Sun is $\approx 15 \pm 5$ km s$^{-1}$, the velocity components of the captured stars with kinematics similar to that of the Sun at the present should lie between the circle of radius of 10 km s$^{-1}$ and that of 20 km s$^{-1}$. Note that only the velocity distributions of the cases $t_f^B(1)$ and $t_f^B(2)$ contain stars with velocities within the mentioned velocity interval.

The total velocities of the same stars, when they were outside the gravitational influence of the perturbing cloud, were of the order of 40 km s$^{-1}$. That is to say that the interaction with the perturbing cloud could reduce the velocities of this star group from 40 to only 15 km s$^{-1}$. The distributions of Figs 12 and 13 correspond to the cloud's equatorial disc of thickness $\Delta H(=50\,\text{pc})$. Since the number of stars between the concentric circles is $\approx 100$ (89 for the case $t_f^B(2)$ and 139 for the case $t_f^B(1)$), the density of G stars with solar kinematics is $\frac{100 f_\star}{\pi a^2 \Delta H} \approx 3 \times 10^{-5}$ stars pc$^{-3}$. Then the number of G stars with solar kinematics in the whole volume of the cloud ($\frac{4}{3}\pi a^3 \sqrt{1-e^2}$) is $\approx 2 \times 10^3$ stars. Certainly, these numbers depends on the values adopted for the parameters, in particular for $V_f^A$. However, they show the explaining capacity of the model. Therefore, the results of the our simulations support the hypothesis delineated above.

### 3.5 Stellar stream formed by stars born in the giant interstellar cloud

In addition to a collector of field stars, the giant interstellar cloud is thought to be a site of star formation. We assume that the new stars are formed during phase B. The disintegration of the cloud, considered in phase C of the model, may be attributed in part to the activity of the stars born in the cloud.

Let us present a simple simulation of phases B and C for the stars born in the cloud. We divide the equatorial disc of the cloud into cells of volume $dV(\Delta x \Delta y \Delta H = 50\,\text{pc} \times 50\,\text{pc} \times 50\,\text{pc})$. We consider that the gas of any element $dV$ is in a circular orbit in the $x$–$y$ plane and gives origin to the formation of a star. The circular orbit of a gas element $dV$ of initial position $(r, \theta)$ is obtained from equations (12) with the initial conditions given by $x_0 = r\cos\theta$, $y_0 = r\sin\theta$, $v_{0x} = -v_c\cos(\pi/2 - \theta)$, $v_{0y} = v_c\sin(\pi/2 - \theta)$, where $v_c = k_B\,r$. Thus, the gas cells rotate in the equatorial plane of the cloud as a rigid body with the angular rotation velocity $\frac{v_c}{r} = k_B$. For the initial conditions of the newborn star, we take random numbers in the ranges $(x_0 - \delta x_0, x_0 + \delta x_0)$ and $(y_0 - \delta y_0, y_0 + \delta y_0)$, for the position, and $(v_{0x} - \delta v_{0x}, v_{0x} + \delta v_{0x})$ and $(v_{0y} - \delta v_{0y}, v_{0y} + \delta v_{0y})$, for the velocity. We adopt $\delta x_0 = \delta y_0 = 25$ pc and $\delta v_{0x} = \delta v_{0y} = 5$ km s$^{-1}$. The number of stars ($=\frac{\pi a^2 \Delta H}{dV}$) adopted for the simulations is certainly arbitrary; and the adoption of uniform number density of stars is not realistic. However, these simulations of illustrative character would permit us to infer the results that the use of more realistic conditions should produce.

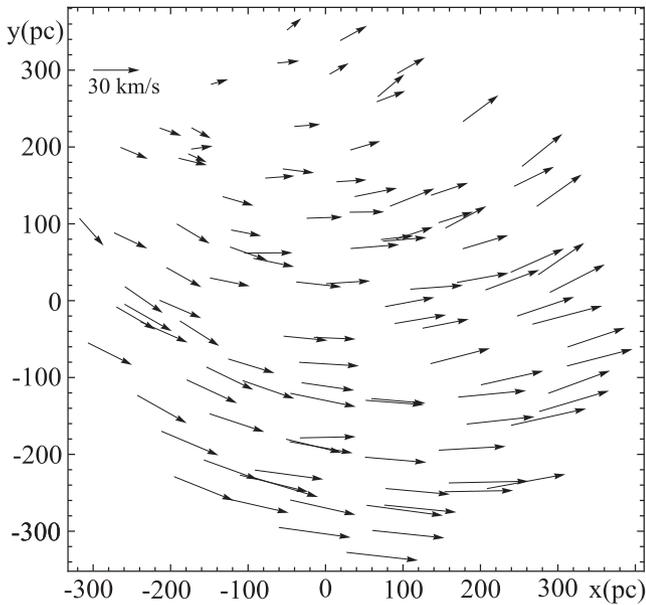

**Figure 14.** Spatial distribution and velocity field, at the end of the final phase (phase C), for the stars born in the cloud.

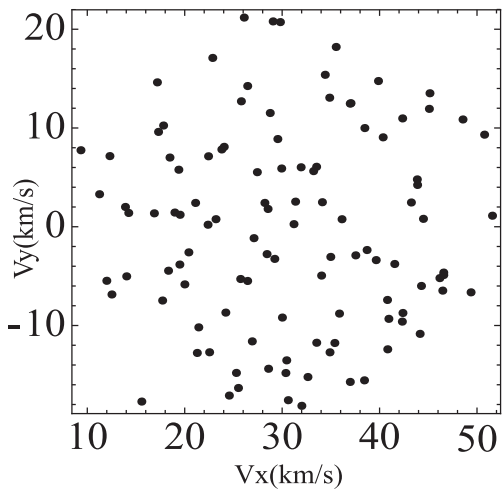

**Figure 15.** Distribution of velocity components corresponding to the velocity field of Fig. 14.

We assume that the stars were born at the same time throughout cloud and have an age, at the end of phase B, between 0 and $t_f^B(i)$ ($= i\frac{T}{4} + nT$, see Section 3.3). Since the absolute values of $t_f^B(i)$ are not necessary for our purposes, we can adopt a fraction of the period as the age of the stars. Since the results obtained using different fractions of the period are very similar, we will take as representative the case $i = 0$. The position and velocity distributions at the end of phase C resulting from our simulation are shown in Figs 14 and 15. The stars born in a certain region of the cloud during a short time interval share close positions in the position and velocity distributions at the end of phase C. Hence, an enhancement of the star formation activity in a relatively small region of the cloud is reflected in the position and velocity distributions as a compact moving group.

## 4 CONCLUSIONS

In the light of our model, giant star–gas complexes of the Galaxy that probably suffered strong accelerations in their genesis would contain an important component of captured field stars. The giant star–gas complexes have typical sizes of ≈650 pc (Alfaro et al. 1992a,b; Efremov 1995, 2010), and in agreement with this, the semimajor axis of the model's cloud is 300 pc. The model's mechanism produces a kinematic selection of the field stars within the cloud, because only the stars with initial velocities close to the direction of the cloud's acceleration are captured. The maximum percentage of capture is obtained with an acceleration that, acting during 10 Myr, produces a cloud's peculiar velocity of ≈30 km s$^{-1}$. Our model shows that, averaged over the whole cloud, an ∼4 per cent of the field stars contained in the original cloud's volume are captured by the interstellar cloud. The subsequent disruption of the cloud's gaseous mass in the last evolutionary stages leaves the stars as moving groups. Therefore, the model can explain the presence of age-homogeneous moving groups as well as of age-heterogeneous ones within the same star complex.

There are two mechanisms that could accumulate interstellar material of a vast region and simultaneously accelerate (or decelerate) it to form a self-gravitating interstellar cloud accelerated or decelerated with respect to the passing field stars. First, according to the classical theory of spiral density waves, a gaseous cloud that enters into a density wave is decelerated and compressed in the associated shock wave. Since the velocity dispersion of the gas is much smaller than that of the stars, the response of the gas to the density waves would be much stronger. This effect could produce a relative deceleration between gas and field stars. However, we should mention that recent simulations of the dynamics of stars and gas around spiral arms show complex results (Baba et al. 2009; Wada, Baba & Saitoh 2011; Grand, Kawata & Croper 2012). From the observational side, direct determinations of distances and proper motions of star-forming regions in spiral arms of our Galaxy with interferometer techniques of very long baseline (Reid et al. 2009 and references therein) show that there are star–gas complexes with peculiar velocities as large as 30 km s$^{-1}$. Secondly, the Milky Way is immersed in a stream of HVCs coming from the Magellanic Clouds. Therefore, these HVCs have been hitting the Galactic disc during a large period of time, and one of the consequences would be the formation of the Galactic warp (Olano 2004; Bobyley 2010; Abedi et al. 2014). The fall of an HVC on the gaseous layer of the Galactic disc can transfer momentum to the gas contained in a large volume without affecting the field stars of the region.

The main aim of the work has been to study the proposed capture mechanism with reasonable astrophysical parameters. The chosen model parameters of 300 pc for the semimajor axis of the cloud and of 14 atoms cm$^{-3}$ for the cloud's maximum density, corresponding to the cloud's mass of $2 \times 10^7$ M$_\odot$, are representative of the sizes and masses of the giant star–gas complexes (Elmegreen & Elmegreen 1983). The velocity of 30 km s$^{-1}$ adopted for the cloud's peculiar velocity due to the acceleration process is approximately the velocity with which the cloud captures the maximum percentage of stars (see Figs 5 and 6). Furthermore, the corresponding velocity distributions (see Figs 9, 11 and 15) cover velocity ranges similar to those of the velocity distribution of single stars of the local moving groups. Indeed, the mean space velocity ($= \sqrt{\overline{U}^2 + \overline{V}^2}$) and the velocity dispersion of the local system of moving groups are ≈25 and ≈± 15 km s$^{-1}$, respectively (see fig. 2 of Montes et al. 2001). The election of the parameter's value for the duration of the acceleration process is, however, more uncertain because of our ignorance of the mechanisms and circumstances involved in the cloud's acceleration (or deceleration). We have adopted a duration of 10 Myr for the acceleration phase, but a difference of $\pm 5$ Myr with respect to the adopted time implies a difference of $\mp 1$ in the capture percentages (see Figs 5 and 6).

Our model is intended as a first step in this line of study of the stellar moving groups. A detailed comparison with the observational data should require to extend this model into a more complex one. This new model should consider that the cloud is in a Galactic orbit and that the motions of the stars associated with the cloud are also affected by the general gravitational field of the Galaxy. In this treatment, the motion equations referred to the cloud's gravity centre contain terms due to the Coriolis force. This force plays an important role, together with the rotation sense of the supercloud, in shaping the moving-group velocity distributions and their vertex deviations (Olano 2001; Bobyley 2004). Then, the values of the model's free parameters can be obtained by fitting the simulation results to the observational data of the moving groups. Another limitation of the model is that we use a rather restrictive criterion of capture, considering as captured only the stars whose orbits remain within the physical contour of the cloud. On the other hand, the approximation of representing the cloud as a homogeneous ellipsoid facilitates the analytic solution, but the real clouds have irregular forms and inhomogeneous mass distributions. The formation and evolution of giant cloud complexes are not well known and the involved physical processes are very complex (e.g. Elmegreen 1979, 1987, 1992). Further theoretical and observational developments in this research field will help to do more realistic simulations of the star capture processes

Finally, we propose that the moving groups of the solar neighbourhood have a common origin in the way considered by our model. The existence of kinematic substructures that characterize the different moving groups would reflex irregularities in the initial spatial distribution of the number density of stars. The model supports also the possibility that the Sun was captured by the same gas–star complex that would have given origin to the moving groups of the solar neighbourhood, and more recently to Gould's belt (Olano 2001). This possibility is of great interest for the studies of natural history of the Earth. Encounters of the Solar system with substructures of the giant cloud could have perturbed recurrently the Oort cometary cloud, causing rains of comets on the Earth (Clube & Napier 1986; Bailey et al. 1990).

## ACKNOWLEDGEMENT

I was benefited from the comments of an anonymous referee.

## REFERENCES

Abedi H., Mateu C., Aguilar L. A., Figueras F., Romero-Gómez M., 2014, MNRAS, 442, 3627
Abramowitz M., Stegun I. A., 1964, Handbook of Mathematical functions. Dover Press, New York
Alfaro E. J., Cabrera-Cano J., Delgado A. J., 1992a, ApJ, 339, 576
Alfaro E. J., Cabrera-Cano J., Delgado A. J., 1992b, ApJ, 386, L47
Allen C. W., 1963, Astrophysical Quantities. Athlone Press, London
Antoja T., Figueras F., Torra J., Valenzuela O., Pichardo B., 2010, in Ulla A., Manteiga M., eds, Lecture Notes and Essays in Astrophysics, Vol. 4, The Origin of Stellar Moving Groups. Tórculo Press, Vigo, p. 13
Antoja T., Figueras F., Romero-Gómez M., Pichardo B., Valenzuela O., Moreno E., 2011, MNRAS, 418, 1413
Asiain R., Figueras F., Torra J., 1999, A&A, 350, 434
Baba J., Asaki Y., Makino J., Miyoshi M., Saitoh T. R., Wada K., 2009, ApJ, 706, 471
Bailey M. E., Clube S. V. M., Napier W. N., 1990, The Origin of Comets. Pergamon, Oxford
Bensby T., Oey M. S., Feltzing S., Gustafsson B., 2007, ApJ, 655, L89
Bhatt H. C., 1989, A&A, 213, 299
Bobyley V. V., 2004, Astron. Lett., 30, 785
Bobyley V. V., 2010, Astron. Lett., 36, 634
Chandrasekhar S., 1942, Principles of Stellar Dynamics. Dover Press, New York
Clube S. V. M., Napier W. N., 1986, in Smoluchowki R., Bahcall J. N., Mattews M. S., eds, The Galaxy and the Solar System. Univ. Arizona Press, Tucson, p. 69
De Silva G. M., Freeman K. C., Bland-Hawthorn J., Asplund M., Bessel M. S., 2007, AJ, 133, 694
De Simone R. S., Wu X., Tremaine S., 2004, MNRAS, 350, 627
Dehnen W., 1998, AJ, 115, 2384
Efremov Yu. N., 1995, AJ, 110, 2757
Efremov Yu. N., 2010, MNRAS, 405, 1531
Eggen O. J., 1965, in Pfenniger D., Bartholdi P., eds, Stars and Stellar Systems, Vol. 5: Galactic Interstellar Medium. Springer, New York, p. 134
Eggen O. J., 1995, AJ, 111, 1615
Eggen O. J., 1996, AJ, 112, 1595
Elmegreen B. G., 1979, ApJ, 231, 372
Elmegreen B. G., 1987, ApJ, 312, 626
Elmegreen B. G., 1992, in Pfenniger D., Bartholdi P., eds, Saas-Fee Advanced Course 21, Swiss Society for Astrophysics and Astronomy, The Galactic Interstellar Medium. Springer-Verlag, Berlin, p. 157
Elmegreen B. G., Elmegreen D. M., 1983, MNRAS, 203, 31
Elmegreen B. G., Elmegreen D. M., 1987, ApJ, 320, 182
Famaey B., Jorissen A., Luri X., Mayor M., Udry S., Dejonghe H., Turon C., 2005, A&A, 430, 165
Famaey B., Siebert A., Jorissen A., 2008, A&A, 483, 453
Fux R., 2001, A&A, 373, 511
Grand R. J. J., Kawata D., Croper M., 2012, MNRAS, 421, 1529
Kapteyn J. C., 1905, Star Streaming. Report of the British Association for the Advancement of Science, South Africa, p. 257, 264
Mihalas D., 1968, Galactic Astronomy. Freeman & Co., San Francisco
Montes D., López-Santiago J., Gálvez M. C., Fernández-Figueroa M. J., DeCastro E., Cornide M., 2001, MNRAS, 328, 45
Nordström B. et al., 2004, A&A, 418, 989
Olano C. A., 2001, AJ, 121, 295
Olano C. A., 2004, A&A, 423, 895
Olano C. A., 2008, A&A, 485, 457
Proctor R. A., 1869, Proc. R. Soc., 18, 169
Quillen A. C., Dougherty J., Bagley M. B., Minchev I., Comparetta J., 2011, MNRAS, 417, 762
Reid M. J. et al., 2009, ApJ, 700, 137
Roberts W. W., 1969, ApJ, 158, 123
Roberts W. W., 1970, in Becker W., Kontopoulos G. I., eds, Proc. IAU Symp. 38, The Spiral Structure of Our Galaxy. Reidel, Dordrecht, p. 415
Wada K., Baba J., Saitoh T. R., 2011, ApJ, 735, 1
Whitman P. G., Matese J. J., Whitmire D. P., 1991, A&A, 245, 75

## APPENDIX A: SOLUTION OF THE MOTION EQUATIONS

(a) *The case of a linear time variation of the supercloud's density*: $\rho(t) = \rho_0(1 + \lambda t)$. The *x*-motion equation (6) is a non-homogeneous equation and in order to obtain its solution, we should first solve the associated homogeneous equation

$$\ddot{x} + k^2(1 + \lambda t)x = 0. \quad (A1)$$

The change of variable $u = u_0(1 + \lambda t)$, where $u_0 = -(\frac{k}{\lambda})^{2/3}$, transforms equation (A1) into an Airy equation: $\ddot{x}(u) - ux(u) = 0$. The solution of the Airy equation is of the form $x(t) = c_1x_1(t) + c_2x_2(t)$, where $x_1(t)$ and $x_2(t)$ are given by the Airy functions of first kind, $Ai(u)$, and of second kind, $Bi(u)$, respectively, and the quantities $c_1$ and $c_2$ are constants. The general solution of equation (6) can be expressed in terms of the solution of equation (A1) and of one particular solution $x_p(t)$ of the nonhomogeneous equation (6) as follows:

$$x(t) = c_1 x_1(t) + c_2 x_2(t) + x_p(t). \tag{A2}$$

The solution $x_p(t)$ can be obtained by means of the method of variation of parameters. Thus, proposing that

$$x_p(t) = c_1(t)x_1(t) + c_2(t)x_2(t), \tag{A3}$$

and imposing arbitrarily the condition $\dot{c}_1(t)x_1(t) + \dot{c}_2(t)x_2(t) = 0$, we can find the functions $c_1(t)$ and $c_2(t)$ that allow $x_p(t)$ to be one solution of the non-homogeneous equation. These functions are $c_1(t) = -\int \frac{f(t)x_2(t)}{W(t)} dt$ and $c_2(t) = \int \frac{f(t)x_1(t)}{W(t)} dt$, where in our case $f(t) = \alpha =$ constant and $W(t) = x_1(t)\dot{x}_2(t) - \dot{x}_1(t)x_2(t)$, the so-called Wronskian determinant. Thus, $W(t) = W\{Ai(u), Bi(u)\}\dot{u}$, where $W\{Ai(u), Bi(u)\} = \pi^{-1}$ (Abramowitz & Stegun 1964). Therefore, $c_1(t) = -\frac{\pi\alpha}{u_0\lambda} \int_0^t Bi(u) dt = -\frac{\pi\alpha}{(u_0\lambda)^2} \int_{u_0}^{u_0(1+\lambda t)} Bi(u) du$, and $c_2(t) = \frac{\pi\alpha}{u_0\lambda} \int_0^t Ai(u) dt = \frac{\pi\alpha}{(u_0\lambda)^2} \int_{u_0}^{u_0(1+\lambda t)} Ai(u) du$. Remembering that $x_1(t) = Ai(u)$ and $x_2(t) = Bi(u)$, and making the corresponding replacements in equation (A3), we obtain

$$x_p(t) = -\frac{\pi\alpha}{(u_0\lambda)^2} Ai(u) \int_{u_0}^{u_0(1+\lambda t)} Bi(u) du$$
$$+ \frac{\pi\alpha}{(u_0\lambda)^2} Bi(u) \int_{u_0}^{u_0(1+\lambda t)} Ai(u) du. \tag{A4}$$

The constants $c_1$ and $c_2$ in equation (A2) are determined from the initial conditions. At $t = 0$, $x_p(0) = 0$, $x_1(0) = Ai(u_0)$, $x_2(t) = Bi(u_0)$, $\dot{x}_1(0) = \dot{A}i(u_0)u_0\lambda$, and $\dot{x}_2(0) = \dot{B}i(u_0)u_0\lambda$. The dot over the Airy functions denotes the derivative with respect to $u$. Hence, $x_0 = x(0) = c_1 Ai(u_0) + c_2 Bi(u_0)$ and $v_{0x} = \dot{x}(0) = c_1 \dot{A}i(u_0)u_0\lambda + c_2 \dot{B}i(u_0)u_0\lambda$. Solving this system of two equations for $c_1$ and $c_2$, we obtain

$$c_1 = \pi x_0 \dot{B}i(u_0) - \frac{\pi v_{0x}}{u_0\lambda} Bi(u_0),$$

$$c_2 = -\pi x_0 \dot{A}i(u_0) + \frac{\pi v_{0x}}{u_0\lambda} Ai(u_0). \tag{A5}$$

Substituting equations (A5) and (A4) into (A2), we have the general solution given by equation (8). The solution of the $y$-motion equation (7) is analogous to that of equation (A1).

(b) *The case of an exponential time variation of the supercloud's density*: $\rho(t) = \rho_0 e^{\mu t}$, where $\rho_0$ and $\mu$ are constants. By similarity with the $x$-motion equation (6), we have

$$\ddot{x} + k^2 e^{\mu t} x = \alpha. \tag{A6}$$

The transformation $u = \frac{2k}{\mu} e^{\mu t}$ converts the homogeneous equation associated with equation (A6) into $\ddot{x}(u) + \frac{1}{u}\dot{x}(u) + x(u) = 0$, the Bessel differential equation of order zero. Its solution is given by $x(u) = c_1 J_0(u) + c_2 Y_0(u)$, where $c_1$ and $c_2$ are arbitrary constants, and $J_0(u)$, $Y_0(u)$ are the so-called Bessel functions of order zero, of the first and second kinds, respectively. If the particular solution of the non-homogeneous equation (A6) is of the form $x_p(t) = c_1(t)J_0(\frac{2k}{|\mu|}e^{\mu t}) + c_2(t)Y_0(\frac{2k}{|\mu|}e^{\mu t})$, we obtain by means of the method of variation of parameters $c_1(t) = -\int \frac{\alpha Y_0(\frac{2k}{|\mu|}e^{\mu t})}{W(t)} dt$ and $c_2(t) = \int \frac{\alpha J_0(\frac{2k}{|\mu|}e^{\mu t})}{W(t)} dt$, where $W(t) = W\{J_0(u), Y_0(u)\}\dot{u} = \frac{\mu}{\pi}$ and $\dot{u} = ke^{\frac{\mu t}{2}}$. Therefore, the general solution of equation (A6) is

$$x(t) = c_1 J_0(u) + c_2 Y_0(u) - \frac{\alpha\pi}{\mu} J_0(u)$$
$$\times \int_0^t Y_0(u) dt + \frac{\alpha\pi}{\mu} Y_0(u) \int_0^t J_0(u) dt, \tag{A7}$$

where $u = \frac{2k}{|\mu|}$. Hence, the velocity is given by

$$\dot{x}(t) = c_1 \dot{J}_0(u)\dot{u} + c_2 \dot{Y}_0(u)\dot{u}$$
$$- \frac{\alpha\pi}{\mu} \dot{J}_0(u)\dot{u} \int_0^t Y_0(u) dt + \frac{\alpha\pi}{\mu} \dot{Y}_0(u)\dot{u} \int_0^t J_0(u) dt. \tag{A8}$$

The dot over the Bessel functions denotes the derivative with respect to the variable $u$. From equations (A7) and (A8) with $t = 0$, and having into account that $\dot{J}_0(u) = -J_1(u)$ and $\dot{Y}_0(u) = -Y_1(u)$ (Abramowitz & Stegun 1964), we obtain a system of two equations relating the initial conditions with the constants $c_1$ and $c_2$, namely $x_0 = x(0) = c_1 J_0(u_0) + c_2 Y_0(u_0)$ and $v_{0x} = \dot{x}(0) = -kc_1 J_1(u_0) - kc_1 Y_1(u_0)$, where $u_0 = \frac{2k}{|\mu|}$. Solving this equation system for $c_1$ and $c_2$ and replacing the values of these constants in equation (A7), we obtain

$$x(t) = \frac{k\pi}{\mu}(J_1(u_0)Y_0(u) - J_0(u)Y_1(u_0))x_0$$
$$- \frac{\pi}{\mu}(J_0(u)Y_0(u) - J_0(u)Y_0(u))v_{0x}$$
$$- \frac{\alpha\pi}{\mu} J_0(u) \int_0^t Y_0(u) dt + \frac{\alpha\pi}{\mu} Y_0(u) \int_0^t J_0(u) dt, \tag{A9}$$

and by similarity

$$y(t) = \frac{k\pi}{\mu}(J_1(u_0)Y_0(u) - J_0(u)Y_1(u_0))y_0$$
$$- \frac{\pi}{\mu}(J_0(u)Y_0(u) - J_0(u)Y_0(u))v_{0y}. \tag{A10}$$

(c) *Comparison of the results obtained from both cases*: we have calculated $\lambda$ from the initial and final density of the cloud in the considered phase (see Table 1). Similarly we obtain that $\mu = \frac{\ln(\frac{\rho_f}{\rho_0})}{t_f}$. Comparing equations (8) and (A9) and, equations (9) and (A10), we note that the values of respective coefficients of the initial conditions and of the respective independent terms are very similar for any fixed value of the parameters. Therefore, the results obtained for both cases are almost coincident.

This paper has been typeset from a T<sub>E</sub>X/LAT<sub>E</sub>X file prepared by the author.